# Deconvolving Instrumental and Intrinsic Broadening in Excited State X-ray Spectroscopies

T.T. Fister[1], G.T. Seidler[1,*], J.J. Rehr[1], J.J. Kas[1], W.T. Elam[2], J.O. Cross[3], K.P. Nagle[1]

1. Physics Department, University of Washington, Seattle, Washington 98105
2. Applied Physics Lab, University of Washington, Seattle, Washington 98105
3. Advanced Photon Source, Argonne National Laboratory, Argonne, IL 60439

*corresponding author: seidler@phys.washington.edu

Intrinsic and experimental mechanisms frequently lead to broadening of spectral features in excited-state spectroscopies. For example, intrinsic broadening occurs in x-ray absorption spectroscopy (XAS) measurements of heavy elements where the core-hole lifetime is very short. On the other hand, nonresonant x-ray Raman scattering (XRS) and other energy loss measurements are more limited by instrumental resolution. Here, we demonstrate that the Richardson-Lucy (RL) iterative algorithm provides a robust method for deconvolving instrumental and intrinsic resolutions from typical XAS and XRS data. For the $K$-edge XAS of Ag, we find nearly complete removal of ~9.3 eV FWHM broadening from the combined effects of the short core-hole lifetime and instrumental resolution. We are also able to remove nearly all instrumental broadening in an XRS measurement of diamond, with the resulting improved spectrum comparing favorably with prior soft x-ray XAS measurements. We present a practical methodology for implementing the RL algorithm to these problems, emphasizing the importance of testing for stability of the deconvolution process against noise amplification, perturbations in the initial spectra, and uncertainties in the core-hole lifetime.

## I. Introduction

In excited-state x-ray spectroscopies the measured spectra are often degraded with respect to the underlying, ideal signal due to the finite core-hole lifetime or practical limits in the experiment's energy resolution. This gives rise to the generic problem wherein a measured spectrum $I$ (referred to as the image) is the result of convolving the true spectrum $O$ (referred to as the object) with a point-spread function $P$ representing the combined effects of intrinsic and experimental broadening, i.e., $I = P \otimes O$, where $\otimes$ represents the convolution. The inverse problem of extracting $O$ from the measured $I$ subject to known $P$ is common to experiments in diverse fields, and various deconvolution procedures have seen extensive theoretical and computational study.[1-30]

A key pragmatic issue is that deconvolution is ill-conditioned both due to the presence of noise and also due to the limited bandwidth of $P$ with respect to the relevant bandwidth of information in $O$. There are two general approaches to handling this issue. First, one may make use of the convolution theorem via Fourier methods, incorporating appropriate filtering or other regularization to stabilize the inversion against noise amplification.[1,5-7,11,28] Second, one may formally model the process of convolution and noise addition, using maximum likelihood principles to determine an optimal deconvolution process subject to the prior information for the particular problem. This typically leads to an iterative algorithm.[11,19,22,27-29] In both cases, the incorporation of prior information is crucial; at a minimum, information about the smoothness of $O$ must be given to stabilize against noise amplification. In the absence of the use of such prior information in the deconvolution process, Loeffen, et al,[18] propose that the Shannon-Hartley theorem[31] from information theory requires that any increase in the bandwidth of the signal (i.e., sharpening of the spectrum) would come at exponential cost in noise amplification.

In x-ray absorption spectroscopy (XAS) measurements of heavy elements the core-hole lifetime $\tau_{core-hole}$ of the K-shell results in a broadening by $\Gamma_{core-hole} = \hbar / \tau_{core-hole}$; this is a monotonically increasing function of the binding energy, reaching values of 10 eV or more for K-shell spectra when $Z > 50$.[32] Then, the measured spectrum is the convolution of the idealized spectrum with a Lorentzian having FWHM of $\Gamma_{core-hole}$. Prior work on deconvolution within the XAS community[1,5,7,13,18,33,34] has focused on Fourier-based techniques. For example, Loeffen, et al,[18] find that one can remove approximately half of lifetime broadening through Weiner filtering. Similarly, Filliponi[7] finds that one can remove two-thirds of the lifetime broadening and greatly improves the behavior at the tails of resonances by using a Gaussian filter and carefully treating endpoint effects and singularities at the edge step. Additional validation for these methods comes through a comparison with resonant inelastic x-ray scattering (RIXS) measurements, from which one can extract idealized XAS spectra independent of core-hole lifetime effects.[35] For example, the ability to resolve quadrupole transitions in the XAS of several rare earth metals using FT-based deconvolution was verified using resonant inelastic x-ray scattering (RIXS).[36] Alternatively, Babanov, et al,[1] find good performance using Tikhonov regularization[29] for an initial estimate of $O$ which is then corrected by a linear, iterative deconvolution algorithm. The results in Babanov, et al.,[1] are promising in that except near the absorption edge, the deconvolved spectra show good agreement with unbroadened calculation of the XAS. Also, Klementev has made good progress on deconvolution in XAFS using a Bayesian approach which incorporates x-ray photoemission results as additional prior information.[13]

The issue of deconvolution also naturally arises in a related, excited state x-ray spectroscopy: the nonresonant inelastic x-ray scattering of hard x-rays from relatively weakly bound shells. This technique is commonly called nonresonant x-ray Raman scattering (XRS). XRS provides a bulk-sensitive alternative to soft x-ray XAS, albeit at the cost of experimental energy resolution, while also accessing additional information through sensitivity to multipole transitions. In typical XRS measurements of light elements, the energy resolution of the measurement is about 1 eV. This is a factor of about 3 to 10 larger than the combined effects of $\Gamma_{core-hole} = \hbar / \tau_{core-hole}$ and the energy resolution in soft x-ray XAS. This broadening is most important in the near-edge region, where one may find important information about local atomic and electronic symmetry in the structure and detailed momentum dependence of the fine structure. With the steady increase in use of the XRS technique in recent years,[37-58] it is important to seek a standard, reliable method for deconvolving the experimental resolution so that XRS and soft x-ray XAS spectra can be compared on similar footing.

We focus here on iterative deconvolution algorithms. Due to their probabilistic nature and the flexibility they offer for the incorporation of prior information, these algorithms have repeatedly demonstrated dramatic improvements over Fourier-based methods for infrared spectroscopies [59,60] and telescope imaging,[28,61] to name two prominent examples. In excited-state x-ray spectroscopies, the experimenter often has significant prior information about the object $O$: the spectra are positive definite and the spectra are relatively smooth wherein the fine structure cannot be too fine (except for pre-edge resonances), the general periodicity in energy of the fine structure changes in a known way with photoelectron kinetic energy, the atomic background is often well-known, etc. The question then arises as to the implementation and efficacy of these modern deconvolution algorithms for excited-state x-ray spectroscopies.

Here, we demonstrate that the Richardson-Lucy (RL) iterative deconvolution algorithm can be used to deconvolve intrinsic, instrumental, or combined broadening from typical XAS or XRS data. The RL approach[19,25] provides a maximum likelihood solution for the unbroadened spectrum subject to Poisson counting statistics,[19,26,28] making it well-suited for x-ray spectroscopies. Various implementations of the RL approach have seen extensive use in deblurring images, including those from the Hubble Space Telescope[62,63] and related problems in astronomy.[17,28] Iterative deconvolution algorithms have previously been used in electron energy loss spectroscopy.[4,9,64] However, to our knowledge, the RL algorithm has not previously been employed in XAS or XRS analysis.

In section II, below, we introduce the RL algorithm and discuss the optimization issues which are specific to excited-state x-ray spectroscopies. In section III, we apply the RL method to the $K$-edge of Ag, which has $\Gamma_{core-hole} = 8.3$ eV. The performance of the RL algorithm for this problem is well beyond the limitations which would be imposed by the Shannon-Hartley theorem, were it applicable, and also significantly better than results to date from Fourier-based deconvolution: starting a few eV after the absorption edge, we find >90% removal of broadening by the combined effects of the core-hole lifetime and the experimental resolution. In section IV, we demonstrate that improved energy resolution can be obtained in XRS measurements by considering the canonical example of the $K$-edge XRS spectrum for diamond. We find full deconvolution of the experimental resolution at least 5 eV away from the edge, with good agreement between the RL-deconvolved XRS spectrum and prior soft x-ray XAS measurements[65] taken with energy resolution comparable to $\Gamma_{core-hole}$. In section V we summarize our findings, discuss their consequences for future experimental practice, and conclude.

**II. The Richardson-Lucy Algorithm**

We model the measured intensity $I(x_i)$ on the discrete grid $\{x_i\}$ by convolving the 'true' spectrum $O(x_i)$ with a point-spread function $P$ representing the combined effects of experimental and intrinsic broadening, and then adding a noise source $N(O, x_i)$, i.e.,

$$I(x_i) = \sum_j P(x_j) O(x_i - x_j) + N(O, x_i) = (P \otimes O)(x_i) + N(O, x_i). \quad (1)$$

The task at hand is to determine the most probable $O$ given $I$, $P$, and the knowledge that $N$ represents independent Poisson statistics at each $x_i$. Bayes' theorem requires

$$p(I|O) = \prod_i \frac{[(P \otimes O)(x_i)]^{I(x_i)} \exp[(P \otimes O)(x_i)]}{I(x_i)}. \quad (2)$$

Following the approach of Shepp and Vardi,[27] the maximum likelihood solution (i.e., when $\partial \ln p(I|O)/\partial O = 0$) is given by the limit of the series of estimators

$$O^{(n+1)} = O^{(n)} \left( P^* \otimes \frac{I}{P \otimes O^{(n)}} \right), \quad (3)$$

where $P^*$ is the grid-reversal of $P$ and where the seed spectrum $O^{(0)} \equiv I$. The $O^{(n)}$ from Eq. 3 are positive definite for positive definite $I$ and are relatively stable with respect to the choice of the point-spread function.[9,11,19]

The RL algorithm in its simplest form (above) will eventually suffer from unphysical levels of noise amplification as the iteration number increases. There are two general responses to this situation. First, one may define a stopping criterion such that iteration of Eq. 3 is halted when the amplification of high frequency noise overtakes the convergence rate of the $O^{(n)}$. Second, the RL algorithm can be regularized for a particular class of problem by use of a constraint or filter function $F$ at each application of Eq. 3, i.e.,

$$O^{(n+1)} = F\left[x_i, O^{(n)}, O^{(n)}\left(P^* \otimes \frac{I}{P \otimes O^{(n)}}\right)\right]. \quad (4)$$

This regularization filter is defined to include a selection of the prior information available about the experiment. In practice, this has included the need for bounds on the percent absorption measured in infrared spectroscopy[60,66] and the rejection of scale-specific fluctuations in the deblurring of astronomical images.[28,67] The flexibility for incorporation of diverse prior information separates the RL and related iterative deconvolution approaches from the direct Fourier-based methods.

We now present a practical implementation of the RL algorithm for deconvolution of XAS and XRS spectra. There are five issues which must be discussed: the deconvolution grid spacing; the incorporation of prior information into a regularizing filter (as in Eq. 4); the use of an appropriate metric for convergence and the completion of iteration; a test of the robustness of the deconvolution against perturbations in the original data; and the determination of the effective energy resolution of the deconvolved spectrum. First, we make use of an energy grid with 0.05 eV spacing; this is chosen to have resolution much finer than any anticipated physics in the problem, thus ensuring that the data grid will not interfere with the process of extracting features of $O$ from $I$. For both XAS and XRS spectra, we use linear interpolation to map the original data onto this finer grid.

Second, there is considerable prior information about the spectra for both XAS and XRS. While a comprehensive treatment of RL deconvolution would make use of all prior information, we find good performance at small computational cost simply by imposing that the spectra should not have features sharp compared to the reconstruction grid spacing. We suppress the generation of high-frequency fine structure in the $O^{(n)}$ by smoothing at each iteration, i.e.,

$$O^{(n+1)} = F \otimes \left(O^{(n)}\left[P^* \otimes \frac{I}{P \otimes O^{(n)}}\right]\right) \quad (5)$$

where $F$ is a Gaussian function with unit integrated amplitude and with a standard deviation $\sigma_{Gauss}$ considerably less than the width of $P$. We use $\sigma_{Gauss}$ between 0.05 and 0.5 eV, depending on the noise level in $I$ and the expected energy scale for the relevant fine structure in $O$. Even with a smoothing filter, convergence often requires several thousand iterations, as we find below. However, this does not provide a practical difficulty: we find typical computation times of about 1 sec per 200 iterations when running interpreted functions in *Mathematica* running on a modern workstation.

Third, we adopt a common metric for the quality of the $O^{(n)}$ by using the $\chi^2$ measure at each iteration,

$$\chi^2_{(n)} = \frac{1}{N} \sum_i^N \frac{\left((P \otimes O^{(n)})(x_i) - I(x_i)\right)^2}{I(x_i)}. \tag{6}$$

For real (noisy) data in the absence of a regularizing filter, noise amplification often results in a local minima in $\chi^2_{(n)}$ and/or $\partial \chi^2_n / \partial n$.[9] Sufficiently large filter widths converge the RL estimate, with larger widths converging faster but at larger asymptotic values of $\chi^2_{(n)}$.

Fourth, it is advisable to examine the robustness of the deconvolution with a few straightforward tests. The effect of additional uncorrelated noise in the initial spectra on the deconvolution is a natural test of stability against perturbations in the initial conditions. The effect of uncertainties in $P$ should also be investigated by repeating the deconvolution process for several examples of physically-allowed $P$, e.g., to allow for uncertainty in $\Gamma_{core-hole}$. Each of these tests is demonstrated in sections III and IV.

Finally, the loss of high-frequency information due to the inclusion of filters, the finite energy spacing in the original data, and the limited bandwidth in $P$ can compromise the ability to reconstruct sharp edges and narrow features. Hence, for each deconvolution problem, we find it necessary to model the deconvolution of a broadened step function and various widths of broadened Lorentzian-shaped resonances to accurately gauge systematic errors near the edge and the amount of remnant broadening after the deconvolution. Examples of this type of modeling will be shown in the next two sections.

## III. Core-Hole Broadening in XAS for the Ag *K*-Edge

We first test RL deconvolution on the Ag *K*-edge XAS, as measured by Kvitky, *et al*[68]; their spectrum in shown in Fig. 1. This spectrum was measured with 0.7 eV spacing in the near-edge region, giving way to constant spacing in photoelectron momentum in the extended regime. For example, the energy spacing between measurements is ~3 eV at 25750 eV and ~ 5 eV near the end of the energy range shown in the figure. For *P*, we use the convolution of a 4 eV-wide Gaussian experimental response function with an 8.32 eV-wide Lorentzian core-hole lifetime (both FWHM), resulting in a 9.25 eV FWHM PSF. The core-hole lifetime is estimated in FEFF using an interpolation of empirical standards.[32,69]

We focus initially on the near-edge region of the spectrum, as this region has the sharpest intrinsic features and hence is most adversely affected by spectral broadening. Our first attempt to deconvolve the Ag *K*-edge XAS spectrum is shown in Fig. 2, where the estimates $O^{(n)}$ for exponentially increasing iteration number *n* are represented by successively darker lines. For reference, the point-spread function is also shown. Clearly, the oscillatory structure in the deconvolved spectra is better resolved and several new features have emerged from shoulders in the original data. Unfortunately, the spectrum shows little convergence in the first 50 eV, likely due to amplification of sharp features. To better control this problem we incorporate a Gaussian filter at each iteration, i.e., as in Eq. 5. The effect of filter width on the convergence rate is shown in Fig. 3, where we see that the RL algorithm converges within 6000 iterations for $\sigma_{Gauss} \geq 0.05$ eV.

In Fig. 4 we further investigate the beneficial stabilization provided by even modest filtering in the deconvolution iteration. In Fig. 4a we show the original data. In Fig. 4b, we show a density-contour plot demonstrating the evolution of $O^{(n)}$ as a function of *n* (from top to bottom) for $\sigma_{Gauss} = 0$. The top-most shading indicates the original data. Note the lack of convergence, indicated by the repeated bifurcations (when moving down the figure) even at high iteration number, together with the angled contours indicating systematic motion in features with increasing *n*. The resulting estimate is shown in Fig. 4c. On the other hand, note the improved convergence in $O^{(n)}$ when using $\sigma_{Gauss} = 0.05$ eV, as we show in Fig. 4d. The contours completely stabilize toward the bottom of the page, with no further bifurcation or drift in spectral features. We show the converged estimate (at $n$ = 20,000) in Fig. 4e. Given these results, we feel confident in the quality of convergence for $\sigma_{Gauss}$ = 0.05 eV. However, it is also important to verify the stability of this solution with respect to perturbation in the initial spectrum.

To this end, we show in Fig. 5 the effect of additive uncorrelated Poisson-distributed noise on the RL deconvolution, again for $\sigma_{Gauss}$ = 0.05 eV. The top curve in the figure is the original data and the successive curves show show 0.1%, 0.3% and 1% noise levels. For the 0.1% additional noise level, the key features in the deconvolved spectra are only weakly affected. Higher noise levels result in strong realization-to-realization variation in the converged RL estimate, $O^{(6000)}$. This can be compensated by a larger $\sigma_{Gauss}$ but at the cost of poorer final energy resolution. Given that the noise from counting statistics in the original data is far below 0.1%, we conclude that the RL inversion with $\sigma_{Gauss}$ = 0.05 eV is stable against the expected experiment-to-experiment variation from Poisson statistics. Increasing $\sigma_{Gauss}$ would stabilize the deconvolution of data with worse counting statistics, at the price of further restricting the degree of deconvolution.

A remaining source of systematic error is the choice of $P$. In the present case, we expect a possible 10% error in our estimate of $\Gamma_{core-hole}$. In Fig. 6, we show the result of RL deconvolution for several $P$ incorporating this error. The range in effective lifetimes, $\Gamma_{eff}$, introduced a natural spread in the height of the deconvolved fine structure, while features near the edge experienced an additional energy shift due to the rapid change in shape at the absorption edge. These modest changes would be unlikely to affect quantitative structural analysis based on the near-edge region.

Restricting high frequency Fourier components in the spectrum due to finite sampling and, especially, filters in the deconvolution algorithm can broaden and introduce systematic errors near the edge step. As shown in Fig. 7, we have simulated attempted recovery of a step function (representing the edge step in $O$) using the above energy spacing and PSF for three widths of Gaussian filters. In each case, various degrees of ringing also appear in the first 5 eV past the edge, introducing a ~10% effect in the intensity near the edge. This phenomenon is localized near the edge, and the first feature at 25525 eV in Fig. 4e is weakly influenced by this effect.

An additional consequence resulting from the loss of high-frequency information is the remnant broadening of sharp features. To model the ability to resolve an individual, sharp feature in $O$, we deconvolve a set of trial Lorentzians with widths ($\Gamma_o$) ranging from 0.1 to 15 eV that had been broadened by the PSF used for the Ag $K$-edge. In Fig. 8 we show the FWHM measured for the original broadened data ($\sigma_I$), the converged RL estimate using $\sigma_{Gauss} = 0.05$ eV ($\sigma_{final}$), and $O$ ($\sigma_O$) as a function of the unbroadened feature's width ($\Gamma_o$). Closer to the edge, the deconvolution is ~80% successful for features with widths near the finest energy feature allowed by the sampling theorem ($2\Delta E = 1.5$ eV) and increasingly successful thereafter. Beyond the first 10 eV, where feature widths are generally 8 eV or higher, we find nearly complete removal of core-hole and experimental broadening.

This conclusion is further validated in Fig. 9. At the top of the figure, we show the original data $I$, $P \otimes O^{(final)}$, and the results of a real space full multiple scattering calculation that includes core-hole lifetime broadening using the FEFF package.[70] Outstanding agreement between $I$ and $P \otimes O^{(final)}$ is evident; if the broadening effects were not substantially removed, then $P \otimes O^{(final)}$ would appear broadened with respect to original data, $I$. In the bottom of the figure we show $O^{(final)}$ and the analogous calculations which exclude core-hole lifetime effects. While the agreement between the broadened theoretical calculation and the original experimental spectrum is good, the comparison between the unbroadened theory and the deconvolved spectrum is much richer. For example, several features have been inferred which existed at best as weak shoulders or inflections in the original spectrum. The disagreement between theory and experiment rests primarily in the relative amplitude of near-edge features; it is unclear if this discrepancy is due to systematic errors (i.e. as in Fig. 7), incomplete information in $I$, or the limitations of the calculation.

RL deconvolution can also be applied beyond the near-edge regime. Recall that XAS oscillations are periodic in photoelectron momentum $k$, not in photoelectron kinetic energy. Consequently, the fine structure has steadily larger energy spacing as photon energy increases past the edge – this is the motivation for the increased spacing between measurements in the original spectrum, mentioned above. However, in the present context, the fact that the original measurement grid in the extended region is not significantly finer than the width of *P* has a negative consequence. The relative paucity of information on this scale results in more difficulties with noise amplification, relative to deconvolution in the near edge region. As a consequence, we use a larger, $\sigma_{Gauss} = 0.5$ eV filter. Using the IFEFFIT package[71] with a fixed edge energy of 25525 eV and a sine window for Fourier transforms, we show $k\chi(k)$ in part (a) of Fig. 10 for the first 670 eV of fine structure using RL estimates up to $n = 130$. The deconvolution dramatically affects the amplitude of the Fourier transform of $k\chi(k)$, as we show in Fig. 10b. In particular, note that the amplitude and intensity of higher coordination shells are greatly enhanced.

**IV. Instrumental Resolution in XRS: C *K*-edge for Diamond**

The choice of incident photon energy in an XRS measurement involves a competition between energy resolution and count rate. As incident photon energy increases the experimental resolution becomes poorer. However, the count rate can be expected to increase because stray absorption effects rapidly decrease, the incoherent portion of the total scattering cross-section increases, and the incident flux from the third generation hard x-ray synchrotrons increases for typical photon energies used in XRS experiments. Balancing these factors typically results in an incident energy of ~10 keV with an overall energy resolution of ~1 eV. Here, we show that reliable knowledge of the point-spread function, together with the good statistics available in contemporary XRS measurements, allow beneficial use of the RL algorithm to improve the effective energy resolution. This then changes the balance of factors in favor of higher incident photon energies for some experiments. Given the recent, dramatic improvements in counting rates for XRS that have resulted from improvements in the efficiency of spherically-bent analyzer crystals[72] and from the development of dedicated multi-element XRS spectrometers,[51,73] we believe that the present results may have significant consequences for future XRS experimental practice.

XRS measurements on the C *K*-edge of diamond were performed at the Advanced Photon Source using the lower energy resolution inelastic x-ray scattering (LERIX) user facility.[41] This previously reported data[41] has ~200,000 counts at each energy point and was taken at a momentum transfer $q$ = 3.9 Å$^{-1}$. The spacing of the energy grid in the original measurement is 0.2 eV. The incident photon energy is ~10 keV with experimental energy resolution FWHM of 1.35 eV. Unlike the previous section, where the accuracy of the PSF was in question, the experimental PSF for an XRS experiment is *measured* at the peak corresponding to elastic scattering. It should be noted that for more tightly bound edges, the monochromator's energy resolution will broaden slightly, which can be a ~10% effect for energy losses beyond 1 keV.

The top two spectra of Fig. 11 are the measured C *K*-edge and the RL-deconvolved spectrum after convergence with a filter having $\sigma_{Gauss}$ = 0.05 eV. The stability of the RL inversion was tested using the previously-described protocol. The RL-deconvolved spectrum compares favorably with 0.3 eV-resolution soft x-ray results taken by Ma *et al*,[65] as we show in the bottom of the figure.

Following the discussion in section III (and Fig. 7), we show in Fig. 12 a systematic investigation of the effect of the sharpness of the underlying edge step on the deconvolution. At the top of the figure, we show absorption edges modeled as step functions convolved with Lorentzians (thus resulting in the usual arctan functional form at the edge), with $\Gamma$ = 0.1, 0.2, and 0.4 eV. Note that $\Gamma_{core-hole}$ is ~0.1 eV for the C 1*s* initial state. This data is then broadened by convolving with *P* from the above XRS experiment (approximately a Gaussian with FWHM of 1.35 eV). Moving down the figure, these spectra are then deconvolved, using Gaussian filters with the indicated characteristics. Hence, the lost intermediate-frequency information in the measurement process again results in a spurious near-edge feature, followed by weak ringing at higher energies.

The case with $\sigma_{Gauss}$ = 0.05 eV reproduces the data processing for the deconvolution in Fig. 11. We therefore conclude that the near-edge peak in our deconvolved diamond spectrum, having apparent agreement with the diamond 1*s* exciton, is largely a spurious consequence of the deconvolution process. The rapid damping of the spurious features in Fig. 12 does, however, reassure us as to the accuracy of the deconvolution for diamond when more than ~3 eV past the edge. In the remainder of the figure we show the benefit of better instrumental energy resolution. An improvement to 0.9 eV experimental resolution for LERIX is available through use of the Si <311> monochromator available at the host beamline. In this case, the integrated intensity in the spurious features is decreased and the deconvolution appears uncorrupted when more than ~2 eV past the edge.

To gauge the degree of deconvolution possible in the diamond data, we performed the analogous test to our treatment of the Ag $K$-edge in Fig. 8. The results, shown in Fig. 13, are similar to the previous case, although full deconvolution now occurs at a much lower feature width due to the narrower PSF. Also, there is not a linear dependence between the FWHM of the broadened signal and the object in this case, since the PSF is not Lorentzian. Here, the deconvolved estimate is not able to fully resolve features with widths below ~2 eV. As this is well above the finest energy feature preserved by the sampling frequency ($2\Delta E = 0.4$ eV), we expect that all features will be fully deconvolved beyond the edge step and the core-exciton at 290 eV.

## V. Discussion and Conclusion

The deconvolution process is inherently ill-conditioned because of information lost to the limited Fourier-spectral range of $P$ and the consequent sensitivity to noise. The importance of this lost information must be evaluated on a case-by-case basis, as must the decision to reintroduce (or at least constrain) the lost information through the details of the deconvolution algorithm. We find here that typical XAS and XRS spectra, especially when away from the absorption edge, are sufficiently bandwidth-limited that accurate and near-complete deconvolution can be obtained while adding little prior information. It is, however, important to understand several limitations of our approach, as they directly affect the practical implementation of the RL algorithm to typical XAS or XRS spectra and, as a result, suggest changes to common data collection practice.

First, there is a natural competition between final energy resolution (i.e., degree of removal of broadening) and noise amplification. We have consequently introduced the methodology described in Section IV for testing the numerical reliability and stability of the deconvolution estimate. We believe that such procedures should be generically applied, for both iterative and Fourier-based methods.

Second, special care is often necessary in validating the accuracy of RL deconvolution very near to the absorption edge. As shown in the case of diamond, the sharp intrinsic edge step cannot be completely recovered. This issue should always be modeled, as shown, for example, in Fig. 12. It would be interesting to add prior information about the shape of the edge in $O$ to the deconvolution process. This information could be incorporated into a more complex filter function. It would also be interesting to separate the treatment of the edge step from the fine structure, in analogy to the approach of Fourier-based approach Filliponi.[7]

Third, while we find good performance with the RL algorithm, which is optimal for Poisson additive noise, there may be cases were additional benefits accrue from the corresponding maximum-likelihood approach for Gaussian additive noise. i.e.,

$$O^{(n+1)} = O^{(n)} + \alpha P^* \otimes (I - P \otimes O^{(n)}). \tag{7}$$

This approach was used by Babanov, et al.[1] Specifically, there may be cases where this algorithm converges more quickly, as the parameter $\alpha$ can be tuned to aid convergence.

As a practical point, it would be interesting to compare how the rate of convergence for the two algorithms is affected for different filter functions incorporating prior information.

Fourth, as pointed out in detail by Klementev's work,[13] it may not be strictly correct to use the same core-hole lifetime at all photoelectron kinetic energies. This can be investigated through use of a shift-variant point-spread function in the RL approach, should it prove important in particular applications.

Finally, the sampling theorem provides an additional mechanism for information loss which we have not addressed. While it is common practice to use relatively wide energy steps $\Delta E$ in measurements of broadened spectra, this results in an unnecessary loss of information (and possibly aliasing) for energy-frequencies above $\pi / \Delta E$. For the Ag $K$-edge data addressed in Section IV, this was not a major concern because the experimenters used $\Delta E$ = 0.75 eV in the near-edge region and (as it turned out) the sharpest features in $O$ were ~ 2 eV wide. However, sampling limitations would have become apparent in the diamond XRS results if the deconvolution at the edge was not contaminated by the ringing phenomenon shown in Fig. 12. The core exciton has a width of ~0.1 eV, while the experimental $\Delta E$ was 0.2 eV. Even in the absence of the issues associated with the very sharp edge-step in diamond, it would have been impossible to recover the core-exciton in $O$ from this experimental data. Consequently, experimenters interested in maintaining the option of deconvolving excited-state x-ray spectra are well advised to use an energy step in the near-edge region which is safely smaller than the finest features which they hope to recover in $O$, rather than safely smaller than the finest features which they immediately observe in $I$.

In conclusion, we find that the Richardson-Lucy (RL) deconvolution algorithm is well-suited to the problem of removing core-hole lifetime effects and instrumental resolution from x-ray absorption fine structure and nonresonant x-ray Raman scattering spectra. We have developed a systematic approach for application of the RL algorithm, including the application of a smoothing filter and several tests for the convergence and accuracy of the deconvolved spectrum. We find near-complete removal of the combined broadening effects of experimental resolution and the core-hole lifetime at the $K$-edge of Ag. We also find greatly improved energy resolution for studies of nonresonant x-ray Raman scattering (XRS), wherein deconvolved XRS spectra for diamond compare favorably with soft x-ray XAS measurements having much finer experimental energy resolution.

**Acknowledgements**

This research was supported by DOE, Basic Energy Science, Office of Science, Contract Nos. DE-FGE03-97ER45628 and W-31-109-ENG-38, ONR Grant No. N00014-05-1-0843, Grant DE-FG03-97ER5623, NIH NCRR BTP Grant RR-01209 and the Summer Research Institute Program at the Pacific Northwest National Lab. The operation of Sector 20 PNC-CAT/XOR is supported by DOE Basic Energy Science,

Office of Science, Contract No. DE-FG03-97ER45629, the University of Washington, and grants from the Natural Sciences and Engineering Research Council of Canada. Use of the Advanced Photon Source was supported by the U.S. Department of Energy, Basic Energy Sciences, Office of Science, under Contract W-31-109-Eng-38. We thank Ed Stern, Micah Prange, Paola D'Angelo, and Robert Pettifer for stimulating discussions and also thank Bud Bridges for supplying the Ag *K*-edge data.

**REFERENCES**

1   Yu. A. Babanov, A.V. Ryazhkin, A.F. Sidorenko et al., Nucl. Instrum. Meth. Phys. Res. A **405**, 378 (1998).
2   P. E. Batson, D. W. Johnson, and J. C. H. Spence, Ultramicroscopy **41**, 137 (1992).
3   W.E. Blass and G.W. Halsey, *Deconvolution of Absorption Spectra*. (Academic Press, New York, 1981).
4   A. F. Carley and R. W. Joyner, Journal of Electron Spectroscopy and Related Phenomena **16**, 1 (1979).
5   J. M. Cole, R. J. Newport, D. T. Bowron et al., Journal of Physics-Condensed Matter **13**, 6659 (2001).
6   R. F. Egerton, B. G. Williams, and T. G. Sparrow, Proceedings of the Royal Society of London Series A **398**, 395 (1985).
7   A. Filipponi, Journal of Physics B **33**, 2835 (2000).
8   B. G. Frederick, G. L. Nyberg, and N. V. Richardson, Journal of Electron Spectroscopy and Related Phenomena **64-5**, 825 (1993).
9   A. Gloter, A. Douiri, M. Tence et al., Ultramicroscopy **96**, 385 (2003).
10  R. Gold, in *AEC Research and Development Report* (Argonne National Laboratory, 1964).
11  P.A. Jansson, *Deconvolution of Images and Spectra*, 2nd ed. (Academic Press, San Diego, 1997).
12  K. Kimoto and Y. Matsui, Journal of Microscopy-Oxford **208**, 224 (2002).
13  K. V. Klementev, Journal of Physics D **34**, 2241 (2001).
14  M. F. Koenig and J. T. Grant, Journal of Electron Spectroscopy and Related Phenomena **33**, 9 (1984).
15  R. Kuzuo and M. Tanaka, Journal of Electron Microscopy **42**, 240 (1993).
16  L. Landweber, American Journal of Mathematics **73** (1951).
17  H. Lanteri, R. Soummer, and C. Aime, Astronomy & Astrophysics Supplement Series **140**, 235 (1999).
18  P. W. Loeffen, R. F. Pettifer, S. Mullender et al., Physical Review B **54**, 14877 (1996).
19  L. B. Lucy, Astronomical Journal **79**, 745 (1974).
20  N. S. McIntyre, A. R. Pratt, H. Piao et al., Applied Surface Science **145**, 156 (1999).
21  M. H. F. Overwijk and D. Reefman, Micron **31**, 325 (2000).
22  E. Picard, Palermo **29**, 79 (1910).
23  S. Prasad, Journal of the Optical Society of America a-Optics Image Science and Vision **19**, 1286 (2002).
24  S. J. Reeves, International Journal of Imaging Systems and Technology **6**, 387 (1995).
25  Richards.Wh, Journal of the Optical Society of America **62**, 55 (1972).
26  W.H. Richardson, Journal of the Optical Society of America **62**, 55 (1972).
27  L.A. Shepp and Y. Vardi, IEEE Transactions on Medical Imaging **MI-1**, 10 (1982).

28    J. L. Starck, E. Pantin, and F. Murtagh, Publications of the Astronomical Society of the Pacific **114**, 1051 (2002).
29    A.N. Tikhonov and V.Ya. Arsenin, *Solution of Ill-Posed Problems*. (John Wiley and Sons, 1977).
30    P.H. van Cittert, Z. Phys. **69** (1931).
31    C.E. Shannon and W. Weaver, *The Mathematical Theory of Communication*. (The University of Illinois Press, Urbana, 1949).
32    O. Keski-Rahkonen and M.O. Krause, Atomic Data and Nuclear Data Tables **14** (1974).
33    Y. A. Babanov, A. V. Ryazhkin, and A. F. Sidorenko, Journal De Physique Iv **7**, 277 (1997).
34    Y. A. Babanov, A. V. Ryazhkin, A. F. Sidorenko et al., Journal of Structural Chemistry **39**, 833 (1998).
35    A. Kotani and S. Shin, Reviews of Modern Physics **73**, 203 (2001).
36    M. H. Krisch, C. C. Kao, F. Sette et al., Physical Review Letters **74**, 4931 (1995).
37    U. Bergmann, P. Glatzel, and S. P. Cramer, Microchemical Journal **71**, 221 (2002).
38    T. T. Fister, G. T. Seidler, C. Hamner et al., Physical Review B **74**, 214117 (2006).
39    D. T. Bowron, M. H. Krisch, A. C. Barnes et al., Physical Review B **62**, R9223 (2000).
40    Y. J. Feng, G. T. Seidler, J. O. Cross et al., Physical Review B **69**, 125402 (2004).
41    T. T. Fister, G. T. Seidler, L. Wharton et al., Review of Scientific Instruments **77**, 063901 (2006).
42    S. Galambosi, M. Knaapila, J. A. Soininen et al., Macromolecules **39**, 9261 (2006).
43    S. Galambosi, J. A. Soininen, K. Hamalainen et al., Physical Review B **6402** (2001).
44    M. L. Gordon, D. Tulumello, G. Cooper et al., Journal of Physical Chemistry A **107**, 8512 (2003).
45    K. Hamalainen, S. Galambosi, J. A. Soininen et al., Physical Review B **65**, 155111 (2002).
46    M. Krisch and F. Sette, Surface Review and Letters **9**, 969 (2002).
47    S. K. Lee, P. J. Eng, H. K. Mao et al., Nature Materials **4**, 851 (2005).
48    W. L. Mao, H. K. Mao, P. J. Eng et al., Science **302**, 425 (2003).
49    W. L. Mao, H. K. Mao, Y. Meng et al., Science **314**, 636 (2006).
50    A. Mattila, J. A. Soininen, S. Galambosi et al., Physical Review Letters **94** (2005).
51    Y. Meng, H. K. Mao, P. J. Eng et al., Nature Materials **3**, 111 (2004).
52    L. A. Naslund, D. C. Edwards, P. Wernet et al., Journal of Physical Chemistry A **109**, 5995 (2005).
53    J. P. Rueff, Y. Joly, F. Bartolome et al., Journal of Physics-Condensed Matter **14**, 11635 (2002).
54    C. Sternemann, J. A. Soininen, S. Huotari et al., Physical Review B **72** (2005).
55    C. Sternemann, J. A. Soininen, M. Volmer et al., Journal of Physics and Chemistry of Solids **66**, 2277 (2005).
56    C. Sternemann, M. Volmer, J. A. Soininen et al., Physical Review B **68** (2003).

57 P. Wernet, D. Nordlund, U. Bergmann et al., Science **304**, 995 (2004).
58 P. Wernet, D. Testemale, J. L. Hazemann et al., Journal of Chemical Physics **123** (2005).
59 P.A. Jansson, *Deconvolution of Images and Spectra*, 2 ed. (Academic Press, San Diego, 1984).
60 P.A. Jansson, R.H. Hunt, and E.K. Pyler, Journal of the Optical Society of America **60** (1968).
61 R.J. Hanisch, R.L. White, and R.L. Gilliland, in *Deconvolution of Images and Spectra*, edited by P.A. Jansson (Academic Press, San Diego, 1984), pp. 310.
62 D. L. Snyder, A. M. Hammoud, and R. L. White, Journal of the Optical Society of America a-Optics Image Science and Vision **10**, 1014 (1993).
63 H. M. Adorf, R. N. Hook, and L. B. Lucy, International Journal of Imaging Systems and Technology **6**, 339 (1995).
64 R. F. Egerton, H. Qian, and M. Malac, Micron **37**, 310 (2006).
65 Y. Ma, N. Wassdahl, P. Skytt et al., Physical Review Letters **69**, 2598 (1992).
66 P.A. Jansson, in *Deconvolutions of Images and Spectra*, edited by P.A. Jansson (Academic Press, San Diego, 1984), pp. 107.
67 J.L. Starck, F. Murtagh, P. Querre et al., Astronomy and Astrophysics **368** (2001).
68 Z. Kvitky, F. Bridges, and G. van Dorssen, Physical Review B **64**, 11 (2001).
69 J.J. Rehr, J.M. de Leon, S.I. Zabinsky et al., Journal of the American Chemical Society **113**, 6 (1991).
70 J.J. Kas, A.P. Sorini, M.P. Prange et al., Physical Review B (in preparation) (2007).
71 B. Ravel and M. Newville, Journal of Synchrotron Radiation **12**, 537 (2005).
72 R. Verbeni, M. Kocsis, S. Huotari et al., Journal of Physics and Chemistry of Solids **66**, 6 (2005).
73 T. T. Fister, G. T. Seidler, L. Wharton et al., Review of Scientific Instruments **77** (2006).

**Figure Captions**

**FIG 1.** Original data taken at the Ag *K*-edge by Kvitky *et al*[68] is shown over its complete energy range.

**FIG 2.** The unfiltered RL estimate is shown for four estimates $O^{(n)}$ such that the darker lines indicate higher iteration number, $n$. For reference, the point spread function (PSF) is also shown.

**FIG 3.** The $\chi^2_{(n)}$ convergence parameter (Eq. 6) for the Ag *K*-edge is shown on a log-log plot for four Gaussian filters. $\chi^2_{(n)}$ was calculated from 25510-25600 eV for the re-convolved estimate with respect to the original data. Note the convergence for $n > 6000$ for $\sigma_{Gauss} \geq 0.05$ eV.

**FIG 4.** The evolution from the original data (a) for the Ag *K*-edge is shown for unfiltered [(b) and (c)] and filtered [(d) and (e)] RL-estimates. The density contour plots [(b) and (d)] represent the logarithmic evolution of the $O^{(n)}$ spectrum as a function of iteration ($n$), with the top representing the original data and the bottom the deconvolved estimate at 20,000 iterations.

**FIG 5.** (a) Uncorrelated Poisson noise was added to the original data at three levels, 0.1%, 0.3%, 1.0%. The curves have been offset for clarity. (b) The RL estimate from three realizations at each noise level for a 0.05 eV width Gaussian filter.

**FIG 6**. Using the converged, 0.05 eV-filtered RL deconvolution estimate, we show the systematic spread resulting from using an effective core-hole width 10% above and below the quoted lifetime.

**FIG 7.** From the top to bottom: a model for an object $O$ representing an unbroadened edge-step; $O$ broadened with the PSF used for the Ag $K$-edge measurements; and the RL estimates for three types of Gaussian filters.

**FIG 8.** Using a Lorentzian with FWHM $\Gamma_o$ for $O$, we plot the final FWHM ($\sigma_{final}$) of the object, the object broadened with the Ag $K$-edge's PSF, and the converged RL estimate using $\sigma_{Gauss}$ = 0.05 eV, as a function of $\Gamma_o$. Note that the amount of remnant spectra broadening ($\sigma_{rem}$) decreases for larger values of $\Gamma_o$.

**FIG 9.** (a) The Ag $K$-edge XAS spectrum, the reconvolution of the RL-estimate $P \otimes O^{(20000)}$, and a broadened theoretical calculation of the Ag $K$-edge XAS (see text for details). The curves have been offset for clarity. (b) The RL estimate $O^{(20000)}$ and the unbroadened theoretical calculation. A Gaussian filter with $\sigma_{Gauss} = 0.05$ eV is used.

**FIG 10.** Using a $\sigma_{Gauss}$ = 0.5 eV filter, we present the extracted fine structure (a) and its Fourier transform (b) for the full range of the data at various stages of deconvolution for the Ag $K$-edge. Higher values of iteration $n$ are denoted by darker lines in each plot.

**FIG 11.** XRS diamond data (top curves) are compared to previous XAS data.[65] The original XRS data is given by the lighter, solid line and is overlaid with the converged, 0.05 eV-filtered RL estimate, given as the darker line. To aid comparison the XRS results have been offset from the XAS data.

**FIG 12.** Systematic error at the edge resulting from the lack of information from higher frequency components is shown for a simulated edge step convolved with a Lorentzian for three lifetimes: 0.1, 0.2, and 0.4 eV. The top curve represents the original simulated edge step, followed by two sets of simulated data with 1.35 and 0.9 eV additional, Gaussian broadening. In each case, three RL estimates using the Gaussian PSF are shown beneath the simulated data corresponding to 0, 0.05, and 0.1 eV width Gaussian filtering.

**FIG 13.** Using a Lorentzian with FWHM $\Gamma_o$ for $O$, we plot the final FWHM ($\sigma_{final}$) of the object, the object broadened with the diamond $K$-edge's PSF, and the converged RL estimate using $\sigma_{Gauss} = 0.05$ eV, as a function of $\Gamma_o$. Note that the amount of remnant spectra broadening ($\sigma_{rem}$) decreases for larger values of $\Gamma_o$.

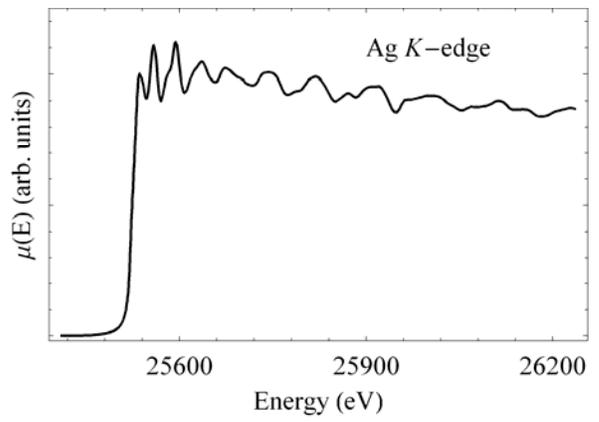

**FIG 1.** T.T. Fister *et al*, "Deconvolving Instrumental …", submitted Phys. Rev. B, 2007.

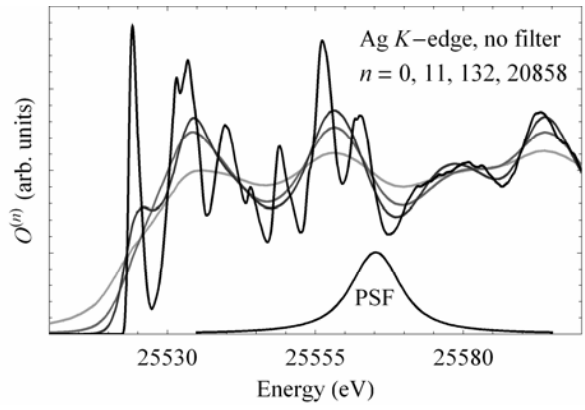

**FIG 2.** T.T. Fister *et al*, "Deconvolving Instrumental …", submitted Phys. Rev. B, 2007.

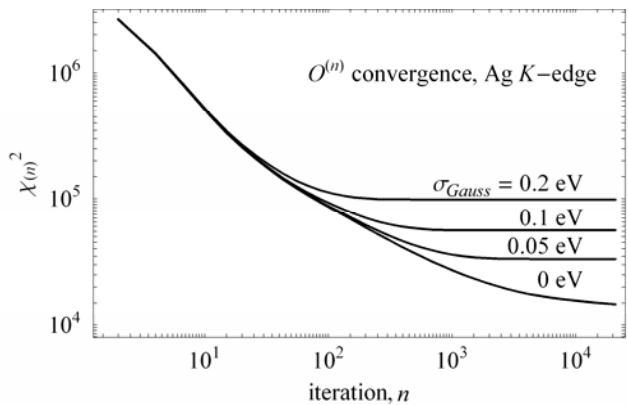

**FIG 3.** T.T. Fister *et al*, "Deconvolving Instrumental …", submitted Phys. Rev. B, 2007.

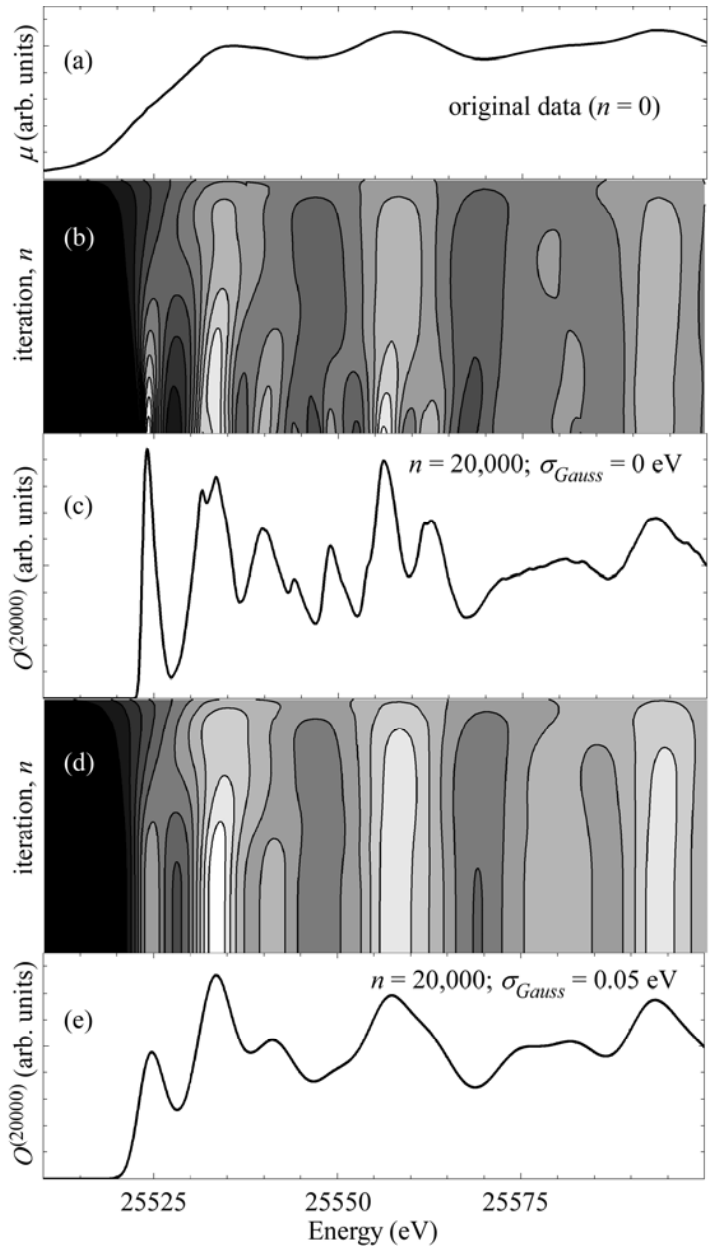

**FIG 4.** T.T. Fister *et al*, "Deconvolving Instrumental …", submitted Phys. Rev. B, 2007.

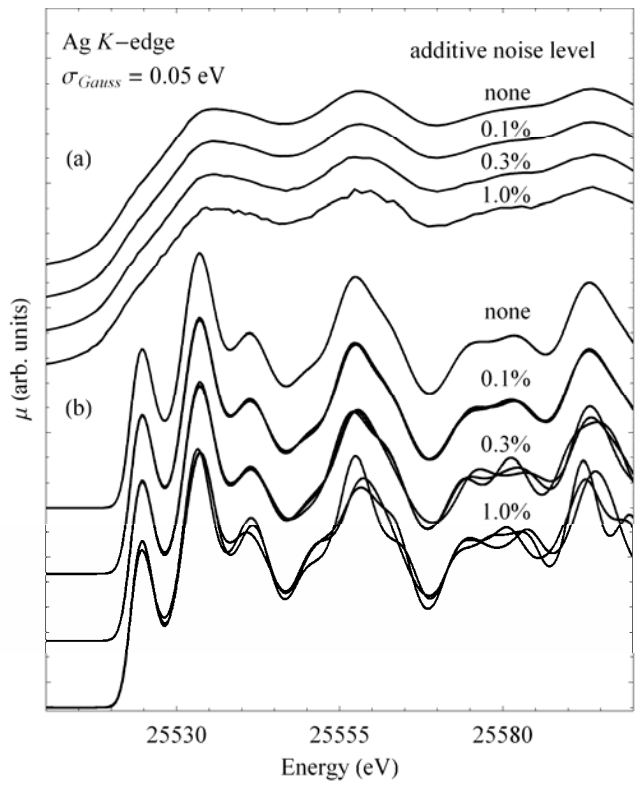

**FIG 5.** T.T. Fister *et al*, "Deconvolving Instrumental …", submitted Phys. Rev. B, 2007.

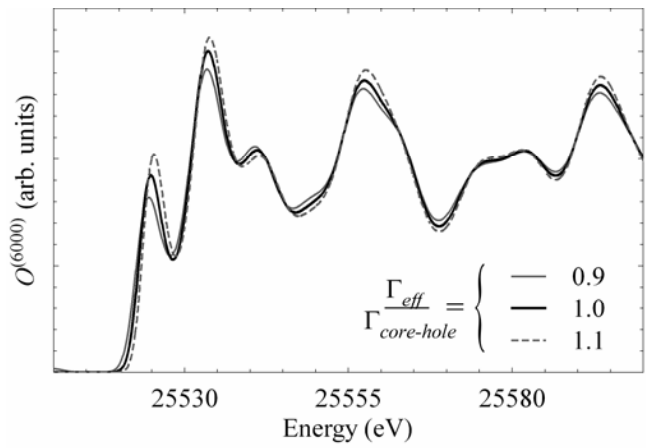

**FIG 6**. T.T. Fister *et al*, "Deconvolving Instrumental …", submitted Phys. Rev. B, 2007.

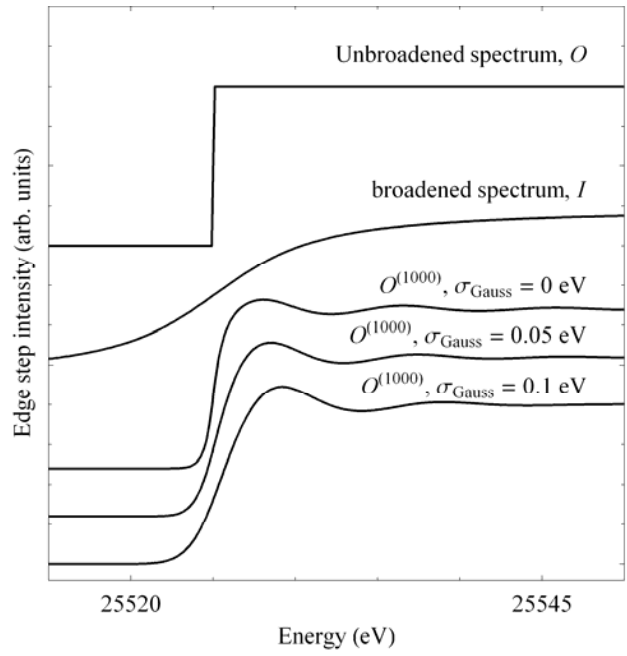

**FIG 7.** T.T. Fister *et al*, "Deconvolving Instrumental …", submitted Phys. Rev. B, 2007.

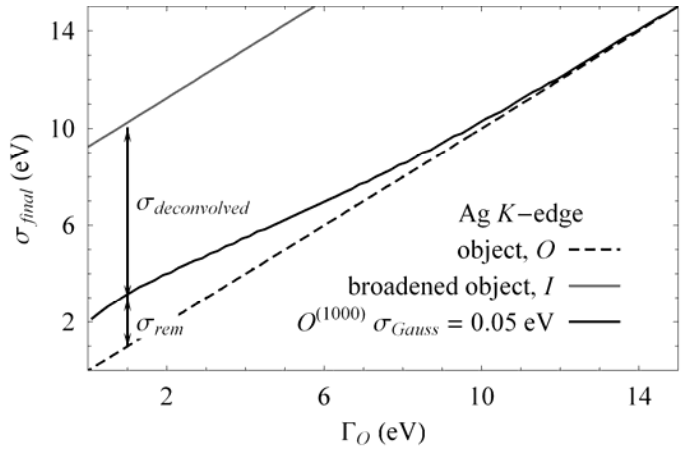

**FIG 8.** T.T. Fister *et al*, "Deconvolving Instrumental …", submitted Phys. Rev. B, 2007.

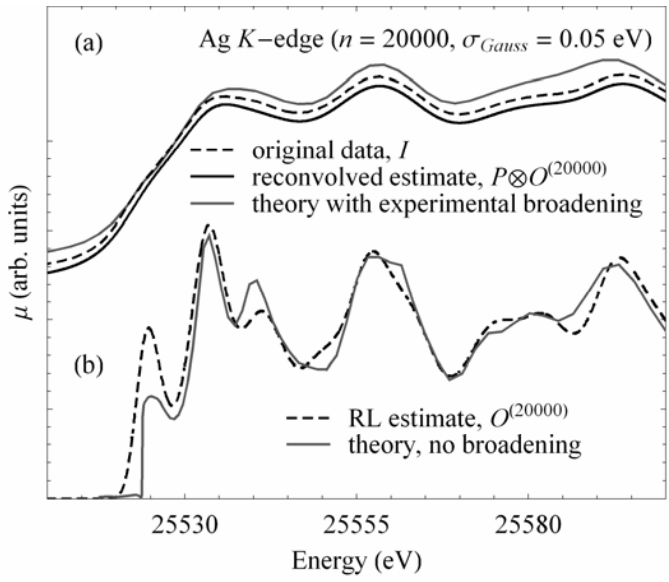

**FIG 9.** T.T. Fister *et al*, "Deconvolving Instrumental …", submitted Phys. Rev. B, 2007.

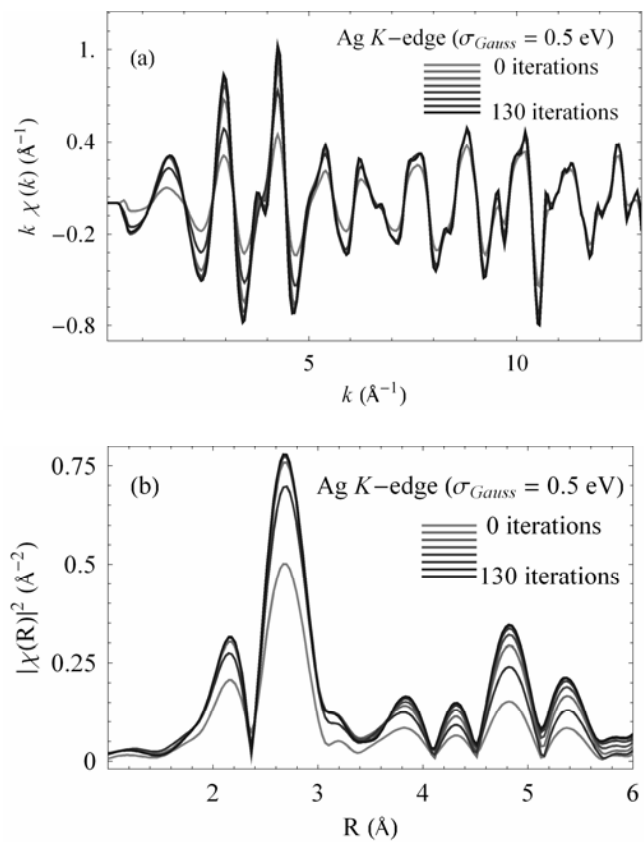

**FIG 10.** T.T. Fister *et al*, "Deconvolving Instrumental …", submitted Phys. Rev. B, 2007.

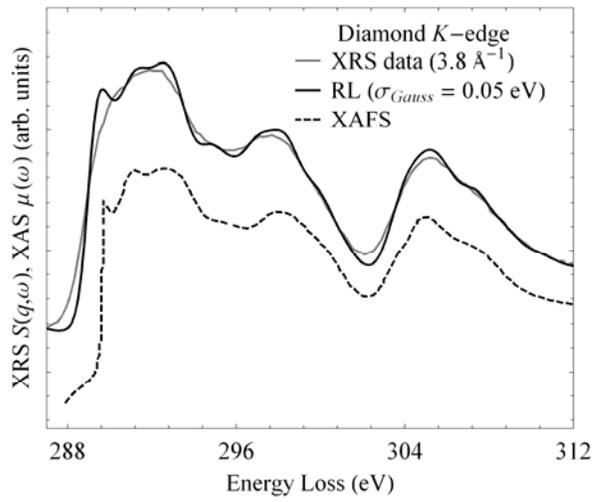

**FIG 11.** T.T. Fister *et al*, "Deconvolving Instrumental …", submitted Phys. Rev. B, 2007.

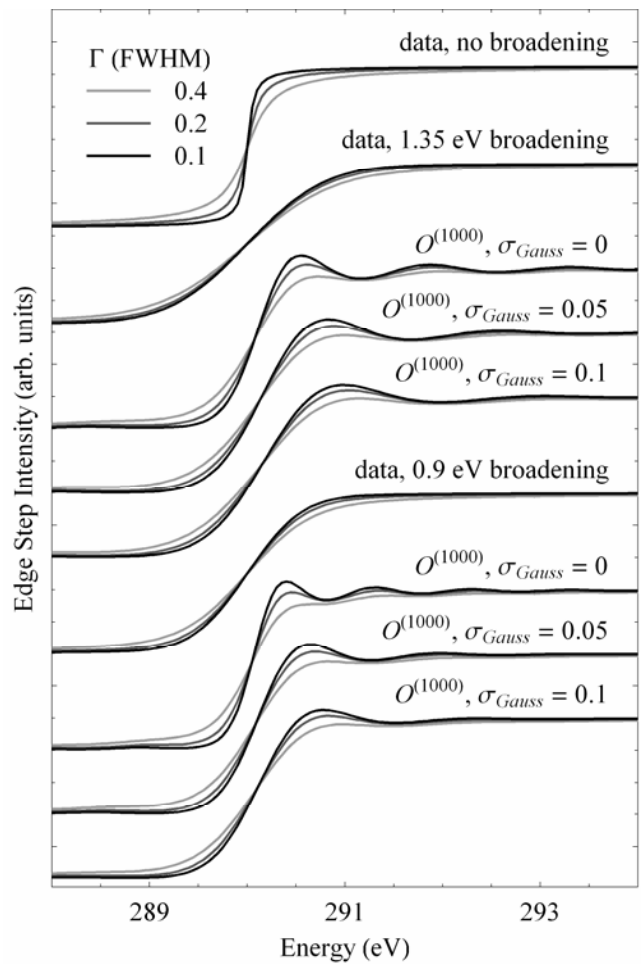

**FIG 12.** T.T. Fister *et al*, "Deconvolving Instrumental …", submitted Phys. Rev. B, 2007.

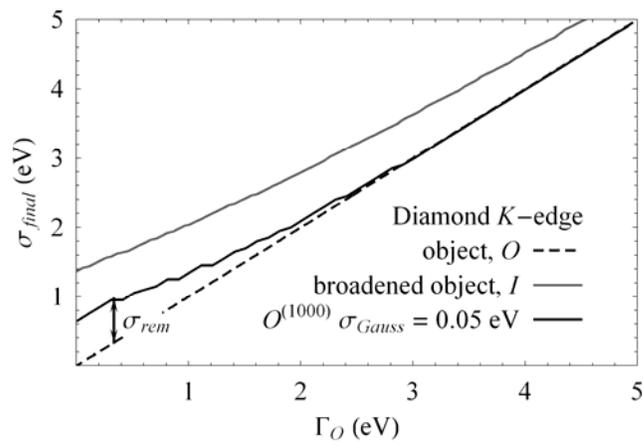

**FIG 13.** T.T. Fister *et al*, "Deconvolving Instrumental …", submitted Phys. Rev. B, 2007.